# Cumulant Approach of Arbitrary Truncated Levy Flight

Dmitry V. Vinogradov[a]

*Econophysics Laboratory of Himex Ltd,*

*30-205 Studencheskaya st., Dzerzhinsk, Nizhny Novgorod oblast, 606016, Russia*

**Abstract.**
The problem of an arbitrary truncated Levy flight description using the method of cumulant approach has been solved. The set of cumulants of the truncated Levy distribution given the assumption of arbitrary truncation has been found. The influence of truncation shape on the truncated Levy flight properties in the Gaussian and the Levy regimes has been investigated.

## 1. Introduction

A truncated Levy flight belongs to the class of discrete random walks with independent, identically distributed increments. The random walk is widely used as a description of physical and non-physical stochastic processes [1].

For example, the physical processes of Brownian motion and diffusion are described by the Gaussian random walk, which is a stochastically self-similar process with a fractal dimension equal to 2 [2].

Stochastically self-similar processes with other fractal dimensions occur every so often in non-physical systems. These are so-called Levy flights [3]. They have infinite variance, and their increments are distributed by the $\alpha$ - stable Levy laws of index $0 < \alpha < 2$ (Levy distributions).

The stochastic processes generated by economical and financial systems are the subject of investigation of the new interdisciplinary approach called econophysics [4,5] and exhibit a number of particular features. The moments of their increments are finite, but the processes themselves are non-Gaussian. Their large-scale fluctuations are close to Brownian motion, whereas the small-scale fluctuations exhibit some of the Levy flight characteristics. A model called truncated Levy flight has been proposed for the description of such stochastic processes for the first time in article [6].

The probability distribution of truncated Levy flight increments is a slightly deformed Levy distribution. This deformation must change the variance of the resulting distribution from the infinite to the finite, and consequently, according to the generalized central limit theorem, the resulting distribution belongs to the Gaussian basin of attraction. The chosen deformation must suppress the "tails" of the Levy distribution and cannot deform the central part. In the pioneering article [6], the abrupt truncation of the Levy distribution tails was used for this purpose.

Instead of the abrupt truncation employed in [6], a smooth exponential regression towards zero was introduced in [7]. This made it possible to derive an analytic expression for stochastic process characteristic function. The scale-invariant non-i.i.d. process, called scale-invariant truncated Levy process as a generalization of the truncated Levy flight was introduced in [8].

---

[a] Corresponding author.  tel: +7(920)0355502, fax: +7(8313) 266092, e-mail: Dmitry.Vinogradov@list.ru

A number of articles [9-11] where other types of truncation are suggested have appeared later. Actually, there are a great number of deformation shapes that are solved in the posed problem. Therefore, a whole class of distributions that can be called "arbitrary truncated Levy distributions" exists.

Notwithstanding the fact that truncated Levy flights have received wide acceptance for the description of stochastic processes of a different nature (see, for example, [12-14]), the special research on the influence of the deformation shape on the stochastic process characteristics has been absent up until now.

The aim of this article is to investigate the deformation shape influence on the arbitrary truncated Levy flight characteristics by the method of cumulant approach.

## 2. Cumulant approach of random walk

The discrete random walk $\{\eta_n\}$,

$$\eta_n = \sum_{i=1}^{n} X_i \tag{1}$$

with independent, identically distributed increments $\{X_i\}$, belongs to the class of non-stationary processes without statistical relationships between increments, and it is non-Gaussian in the general case. This stochastic process is completely described [15] by the one-dimensional probability distribution function (pdf) $W(x,n)$, where $n$ is a step number. It should be noted that $W(x,1) \equiv P(x)$ is the pdf of increments $\{X_i\}$. The one-dimensional characteristic function $\theta(q,n)$,

$$\theta(q,n) = \int_{-\infty}^{+\infty} W(x,n) e^{iqx} dx, \tag{2}$$

which is a Fourier transform of the one-dimensional pdf $W(x,n)$ [16], is the second comprehensive description of the stochastic process (1).

The method of cumulant approach is successfully used in statistical radio-physics [15] for studying non-Gaussian random processes, which include the case under consideration. The random process (1), in the framework of the cumulant approach, is unambiguously and exhaustively described by the infinite set of cumulant functions $\tilde{\kappa}_j(n) \quad j = 1, 2, 3\ldots$, where

$$\theta(q,n) = \exp\left[\sum_{j=1}^{\infty} \frac{\tilde{\kappa}_j(n)}{j!} (iq)^j\right]. \tag{3}$$

The set of cumulant functions is a fundamental [15], and consequently the cumulant approach is one of the most simple and the most powerful methods for studying non-Gaussian stochastic processes. Indeed, by virtue of the absence of statistical relationships between increments, in the frame of the cumulant approach, the stochastic process (1) is described [15] by the cumulant functions with linear dependence on the step number:

$$\tilde{\kappa}_j(n) = n \cdot \kappa_j, \tag{4}$$

where $\kappa_j \equiv \tilde{\kappa}_j(1)$ is the set of cumulants, which determine the pdf of increments $P(x)$.

As mentioned above, cumulant functions exhaustively and unambiguously describe a random process and contain all characteristics of the process under investigation. For example, for the second cumulant function $\tilde{\kappa}_2(n) \equiv D(n)$, which is the variance of process, one obtains the well known diffusive law [2] from expression (4):

$$D(n) = \mathcal{D} \cdot n \tag{5}$$

where the diffusive constant $\mathcal{D} = \kappa_2 = \sigma^2$ is the variance of increments.

The shape of the one-dimensional pdf is defined by the cumulants as well, or, more precisely, by the dimensionless cumulant coefficients (standardized cumulants) [15] $\lambda_j = \kappa_j / \sigma^j$. A normal distribution is characterized by two cumulants of the first order (the mean and the variance). The third and the fourth cumulant coefficients $\lambda_3$, $\lambda_4$ are the skewness coefficient and the kurtosis coefficient, respectively. The higher order cumulant coefficients describe more complicated differences between this distribution and the normal distribution.

The shape of the stochastic process $n$ th step one-dimensional pdf $W(x,n)$ is described by the high cumulant coefficients $\tilde{\lambda}_j(n) = \tilde{\kappa}_j(n) / \tilde{\kappa}_2^{j/2}(n)$, which can be obtained from expression (4):

$$\tilde{\lambda}_j(n) = \frac{\lambda_j}{n^{j/2-1}}, \tag{6}$$

where $\lambda_j$ is the $j$th order cumulant coefficient of the increment pdf. The faster the high cumulant coefficients $\tilde{\lambda}_j(n)$ tend to zero, the high their orders. When the number of steps tends to infinity, the high order cumulants vanish, and the one-dimensional pdf $W(x,n)$ is described by the first two cumulants only or becomes Gaussian. Expression (6) is none other than the cumulant approach of the central limit theorem.

Hence, it is clear that the random walk process with independent identically distributed increments (1) is described by the cumulant functions (4) and completely determined by the increment cumulants.

## 3. Cumulants of arbitrary truncated Levy flight

It is necessary to find the cumulants of a truncated Levy distribution, which is the increment distribution of the stochastic process for the investigation of deformation shape influence on Levy flight properties. Only the symmetrical case will be studied here. Let us assume an arbitrary truncated Levy distribution as a product of two functions:

$$P(x) = C P_L(x) g(x), \tag{7}$$

where $P_L(x)$ is the symmetrical not drifted $\alpha$- stable Levy distribution [17,18], $g(x)$ is the symmetrical deformation function, and $C$ is a normalizing constant. Let us assume that the deformation function equals a unit at the origin $g(0)=1$, and it tends to zero simultaneously with all of their derivatives faster than every power of $1/|x|$ when the argument tends to the infinity $|x| \to \infty$. Let us also assume that the characteristic spatial scale $l$ of the deformation function $g(x)$ is many times greater than the characteristic spatial scale $\gamma$ of the initial Levy distribution:

$$l \gg \gamma. \tag{8}$$

To find of the set of distribution cumulants $\kappa_j$, let us use the cumulant expression as a function of moments $m_j$ [19]:

$$\kappa_j = j! \sum_{i=1}^{j} \sum \left(\frac{m_{p_1}}{p_1!}\right)^{\pi_1} \dots \left(\frac{m_{p_i}}{p_i!}\right)^{\pi_i} \frac{(-1)^{\rho-1}(\rho-1)!}{\pi_1! \dots \pi_i!}, \tag{9}$$

where the second summation is taken over all positive $\pi$ and $\rho$, which obey the conditions:

$$p_1 \pi_1 + p_2 \pi_2 + \dots + p_i \pi_i = j \tag{10}$$

and

$$\pi_1 + \pi_2 + \dots \pi_i = \rho. \tag{11}$$

The odd cumulants equal zero due to the symmetry of distribution (7), and the lower cumulants are

$$\kappa_2 = m_2$$
$$\kappa_4 = m_4 - 3m_2^2 \qquad (12)$$
$$\kappa_6 = m_6 - 15 m_2 m_4 + 30 m_2^3$$

The moments of distribution (7), in turn, can be found from the value of characteristic function $\theta(q)$ derivatives at the origin [16]:

$$m_j = (-i)^j \left[ \frac{d^j \theta(q)}{dq^j} \right]_{q=0}. \qquad (13)$$

where $\theta(q)$ is the Fourier transform of the pdf (7).

Because the truncated Levy distribution (7) is expressed as a product of two functions, based on the properties of Fourier transforms [20], the characteristic function $\theta(q)$ is the convolution of their Fourier transforms:

$$\theta(q) = \frac{C}{2\pi} \cdot \int_{-\infty}^{+\infty} \theta_L(q-\tau) \cdot G(\tau) d\tau. \qquad (14)$$

where $G(q)$ is a Fourier transform of the deformation function $g(x)$ and

$$\theta_L(q) = \exp\left(-\gamma^\alpha |q|^\alpha\right), \qquad (15)$$

is a characteristic function of the symmetric non-shifted $\alpha$-stable Levy distribution (see, for example [16-18]), where $\alpha$ is the index of stability and $\gamma$ is the spatial scale.

Based on the properties of the characteristic function [16], namely, $\theta(0) = 1$ and $\theta(-q) = \theta^*(q)$, where the sign $*$ means the complex conjugate value, and the symmetry of the distribution, one obtains from expressions (7), (13)-(15):

$$m_j = (-i)^j \frac{\int_{-\infty}^{+\infty} \theta_L^{(j)}(q) \cdot G(q) dq}{\int_{-\infty}^{+\infty} \theta_L(q) \cdot G(q) dq}. \qquad (16)$$

Furthermore, one can use the presence of a small parameter $\varepsilon = (\gamma/l)^\alpha \ll 1$, namely, the ratio of the spatial scales of the Levy distribution and the deformation function, and find the moments of the truncated Levy distribution similar to an asymptotic expansion of the small parameter $\varepsilon$.

Let us use the Laplace method (see for example [21]). The Fourier transform $G(q)$ of the deformation function is concentrated in the small neighborhood of the origin in the form of sharp maxima and vanishes away from this region. The sharper this maxima is, the smaller the parameter $\varepsilon$. On the other hand, the value of the characteristic function of the Levy distribution $\theta_L(q)$ undergoes small changes in this neighborhood. As a result, the exact value of function $\theta_L(q)$ can be replaced by an asymptotic expansion in the neighborhood of the origin, and the solution must be found as a power of the small parameter $\varepsilon$.

Let us enter dimensionless coordinates $\varsigma = l \cdot q$, $\xi = x/l$ and expand the characteristic function $\theta_L(\varsigma)$ with the asymptotic series of powers of the small parameter $\varepsilon$ to find

$$\theta_L(\varsigma) = \exp\left(-\varepsilon |\varsigma|^\alpha\right) \approx 1 - \varepsilon |\varsigma|^\alpha + \varepsilon^2 \frac{|\varsigma|^{2\alpha}}{2} - \ldots \qquad (17)$$

One changes the variables in expression (16) to dimensionless ones and replaces the exact value of the characteristic function by an approximate one (17). Retaining terms of order up to $\varepsilon^1$ inclusive, one rewrites expression (16) for the moments of the truncated Levy distribution as

$$m_j = -(-i)^j l^j \varepsilon \cdot \alpha(\alpha-1)\ldots(\alpha-j+1) \int_{-\infty}^{\infty} |\varsigma|^{\alpha-j} G(\varsigma) d\varsigma. \tag{18}$$

Taking into account that the inverse Fourier transform $\mathcal{F}^{-1}$ of a power function [22] is

$$\mathcal{F}^{-1}\left[|\varsigma|^\beta\right] = -\frac{1}{\pi} \sin\frac{\pi\beta}{2} \Gamma(\beta+1)|\xi|^{-\beta-1}, \tag{19}$$

where $\Gamma(x)$ is the Gamma function, and converting from the frequency domain to the time domain expression (18), the result is

$$m_j = l^j \varepsilon \frac{\Gamma(\alpha+1)}{\pi} \sin\frac{\pi\alpha}{2} \int_{-\infty}^{\infty} |\xi|^{j-1-\alpha} g(\xi) d\xi. \tag{20}$$

Expression (20) is true for all moments of an arbitrary truncated Levy distribution. It should be noted that orders of the truncated Levy distribution moment magnitudes are $m_j \sim l^j \varepsilon$. Therefore, retaining terms of order up to $\varepsilon^1$, one obtains from expressions (9)-(11)

$$\kappa_j = m_j = l^{j-\alpha} \gamma^\alpha A(\alpha) \cdot \mu_j(\alpha), \tag{21}$$

where

$$A(\alpha) = \frac{2}{\pi} \Gamma(\alpha+1) \sin\frac{\pi\alpha}{2} \tag{22}$$

and

$$\mu_j(\alpha) = \int_0^{\infty} \xi^{j-1-\alpha} g(\xi) d\xi, \tag{23}$$

where the function $\mu_j(\alpha)$ describes the truncation shape influence on the cumulants. It should be noted that the function $\mu_j(\alpha)$ is the Mellin transform of the deformation function $g(\xi)$ [20].

Expressions (21)-(23) are simplified if the index of stability is $\alpha=1$ (truncated Cauchy distribution). In this case, the values of influence function $\mu_j(1)$ coincide with the deformation function moments of corresponding orders $\mu_j(1) = M_{j-2}$, and one obtains the simple expression for cumulants:

$$\kappa_j = l^{j-1} \gamma \frac{2 M_{j-2}}{\pi}. \tag{24}$$

As mentioned above, the high cumulant coefficients describe the differences between this distribution and the normal distribution. Based on expressions (21)-(23), one obtains the high cumulant coefficients of truncated Levy distribution:

$$\lambda_j = \left(\frac{l}{\gamma}\right)^{\alpha(j-2)/2} \frac{\mu_j(\alpha)}{A(\alpha) \mu_2^{j/2}(\alpha)}. \tag{25}$$

Given the results obtained, the distinguishing feature of an arbitrary truncated Levy distribution, from the cumulant approach point of view, is the specified orders of cumulant coefficients magnitudes: $\lambda_j \sim \varepsilon^{1-j/2}$, (for example $\lambda_4 \sim \varepsilon^{-1}$, $\lambda_6 \sim \varepsilon^{-2}$ etc.). This dependence is a "visiting card" of a truncated distribution, and any probability distribution belongs to the class of truncated Levy distributions if their cumulant coefficients fulfill these requirements.

It follows from the results derived that the cumulants of the truncated Levy flight (21) are directly dependent on the spatial scale $l$ of the deformation function as well as on the ratio of spatial scales $\varepsilon = (\gamma/l)^\alpha$, namely, $\kappa_j \sim l^j \varepsilon$. It should be noted that by virtue of the small size of this ratio $(\gamma/l)^\alpha \ll 1$ the cumulant dependence $\kappa_j(\alpha) \sim (\gamma/l)^\alpha$ upon the initial Levy distribution stability index $\alpha$ is strong, whereas expression (22) describes the weak dependence upon the stability index $\alpha$. The cumulant dependence upon the deformation function shape is described by expression (23).

## 4. Examples of truncated Levy distributions

Let us consider several examples of deformation functions as an illustration of the results obtained above.

*1. Mantegna – Stanley truncation.* The truncated Levy flight was proposed in article [6] for the first time, and the abrupt truncation of distribution "tails" was used. The deformation function corresponding to the truncation used,

$$g_{ms}(\xi) = \begin{cases} 1, & |\xi| \leq 1 \\ 0, & |\xi| > 1 \end{cases}. \tag{26}$$

generates the influence functions $\mu_j(\alpha)$

$$\mu_j(\alpha) = \frac{1}{j-\alpha}. \tag{27}$$

All cumulants of the given truncated distribution can be obtained from expressions (21)-(23), and, for example, the variance is

$$\sigma^2 = l^{2-\alpha} \gamma^\alpha A(\alpha) \cdot \frac{1}{2-\alpha}. \tag{28}$$

This result coincides with the result for the variance derived in article [6]. However, it should be noted that the variance was obtained in [6] for the limited domain of the stability index $1 \leq \alpha < 2$, whereas now it is obtained for the whole domain of stability index $0 < \alpha \leq 2$.

In addition, the kurtosis coefficient is a sufficiently important characteristic of a truncated Levy distribution, and the fourth cumulant coefficient for the Mantegna-Stanley truncation is

$$\lambda_4 = \left(\frac{l}{\gamma}\right)^\alpha \frac{1}{A(\alpha)} \frac{(2-\alpha)^2}{(4-\alpha)}. \tag{29}$$

*2. Exponential truncation.* Another significant example is exponential suppression, and the appropriate deformation function is

$$g_e(\xi) = \exp(-|\xi|), \tag{30}$$

which generates the influence function

$$\mu_j(\alpha) = \Gamma(j-\alpha). \tag{31}$$

Correspondingly, in the case of exponential truncation the variance is

$$\sigma^2 = l^{2-\alpha} \gamma^\alpha A(\alpha) \cdot \Gamma(2-\alpha), \tag{32}$$

and the fourth cumulant coefficient is

$$\lambda_4 = \left(\frac{l}{\gamma}\right)^\alpha \frac{1}{A(\alpha)} \frac{(2-\alpha)(3-\alpha)}{\Gamma(2-\alpha)}. \tag{33}$$

*3. Power-exponential truncation.* A deformation function with power-exponential suppression can be proposed

$$g_s(\xi) = \exp(-|\xi|^h), \tag{34}$$

possessing the parameter $h$, which changes the shape of the truncation and the amount of Levy distribution "tail" suppression from the total suppression when $h \to \infty$ to the total absence of suppression when $h = 0$. The corresponding influence function is

$$\mu_j(\alpha) = \frac{1}{h}\Gamma\left(\frac{j-\alpha}{h}\right). \tag{35}$$

It should be noted that if $h \to \infty$ the influence function of power-exponential truncation tends to the one of the Mantegna-Stanley truncation.

The variance of power-exponential truncated distribution is

$$\sigma^2 = l^{2-\alpha}\gamma^\alpha A(\alpha)\cdot\frac{\Gamma((2-\alpha)/h)}{h}, \tag{36}$$

and the fourth cumulant coefficient is

$$\lambda_4 = \left(\frac{l}{\gamma}\right)^\alpha \frac{h}{A(\alpha)}\frac{\Gamma((4-\alpha)/h)}{\Gamma^2((2-\alpha)/h)}. \tag{37}$$

## 5. Behaviors of truncated Levy flight

It is known [4-6] that behaviors of truncated Levy flight fluctuations depend on their scale $N$. When the large-scale fluctuations of a process have the nature of Brownian motion, it is called *the Gaussian regime* of truncated Levy flight [4,6]. The small-scale fluctuations have some properties of Levy flight fluctuations, and it is called *the Levy regime*.

*Gaussian regime.* From the cumulant approach point of view the Gaussian regime occurs when the high cumulant coefficients of the truncated Levy flight one-dimensional probability distribution $W(x,n)$ can be neglected. The process in the Gaussian regime is described by the diffusion law (5), though the diffusion coefficient is the variance of the increment distribution. The variance of the truncated Levy distribution, which is the increment distribution of the truncated Levy flight, is evaluated with expression (21) and in the particular cases by expressions (28), (32), (36).

As follows from (6), the Gaussian regime comes when the characteristic spatial scale $N_G$ of the fluctuations exceeds the fourth cumulant coefficient (kurtosis coefficient) of the increment distribution $N_G \gg \lambda_4$, and for the truncated Levy flight one obtains $N_G \gg \lambda_4 \sim (l/\gamma)^\alpha$ or, more precisely,

$$N_G \gg \lambda_4 = \left(\frac{l}{\gamma}\right)^\alpha \frac{\mu_4(\alpha)}{A(\alpha)\mu_2^2(\alpha)}, \tag{38}$$

From the cumulant approach point of view the Gaussian regime is completely described by the second and the fourth cumulants (the variance and the kurtosis) of the increment distribution. On the other hand, these cumulants depend on the Levy distribution shape of the truncation (21) –(23), and the shape of the truncation influences both the diffusion coefficient and the Gaussian regime condition (38).

For example, for the stochastic process of truncated Cauchy flight (the stability index $\alpha = 1$) and the Mantegna-Stanley truncation (26) one obtains the diffusion coefficient $\mathcal{D} = 2l\gamma/\pi$ and the Gaussian regime condition $N_G \gg \pi l/6\gamma$.

For the exponential truncation (30) one obtains the same diffusion coefficient $\mathcal{D} = 2l\gamma/\pi$ but another Gaussian regime condition. The Gaussian regime condition for the exponential truncation differs from the one for the Mantegna-Stanley truncation by a factor of six- $N_G \gg \pi l/\gamma$.

For the power-exponential truncation (34) and the value of parameter $h=1/2$, the diffusion coefficient is greater than the one for the Mantegna-Stanley truncation by a factor of two-- $\mathcal{D} = 4l\gamma/\pi$ --and the Gaussian regime conduction increases by 180 times -- $N_G \gg 30\pi l/\gamma$.

*Levy regime.* In the case of truncated Levy flight small-scale fluctuations, when the characteristic spatial scale of fluctuations obeys the condition $N \leq (l/\gamma)^\alpha$, the high cumulants of the pdf cannot be neglected. In this case, one obtains the truncated Levy flight one-dimensional characteristic function from expressions (30), (4), and (21), taking into account the condition (8):

$$\theta(q,n,\gamma) = \exp\left[\sum_{j=1}^{\infty} \frac{n \cdot \gamma^\alpha l^{j-\alpha} A(\alpha) \cdot \mu_j(\alpha)}{j!}(iq)^j\right]. \tag{39}$$

Due to the fact that the dependence $\kappa_j \sim l^{j-\alpha}\gamma^\alpha$ is typical for the truncated Levy distribution cumulants, one can obtain the following equality from expression (39):

$$\theta(q,n,\gamma) = \theta(q,1,\gamma \cdot n^{1/\alpha}). \tag{40}$$

The physical meaning of equality (40) is the following. The *n*th step one-dimensional probability distribution of the Levy flight coincides with the distribution that can be obtained from the initial truncated Levy distribution by changing its spatial scale from the value $\gamma$ to the value $\gamma_{eff} = \gamma \cdot n^{1/\alpha}$ keeping with the spatial scale of the deformation function $l$. This statement is correct under the condition

$$\gamma_{eff} = n^{1/\alpha}\gamma \ll l \tag{41}$$

when the cumulants of the truncated Levy flight with the spatial scale $\gamma_{eff}$ obey expressions (21)-(23).

The known properties of the truncated Levy flight returns [4-6], namely, the return dependence on the step number, simply follows from equality (40). Actually, one finds from (40) that

$$W(0,n,\gamma) = W\left(0,1,\gamma \cdot n^{\frac{1}{\alpha}}\right) \tag{42}$$

Under condition (41), the influence of the deformation function on the return is the second order of the smallness value, and the return equals the one of an undisturbed Levy flight:

$$W(0,n,\gamma) = P_L(0,\gamma \cdot n^{1/\alpha}) = \frac{\Gamma(1/\alpha)}{\pi\alpha\gamma n^{1/\alpha}}. \tag{43}$$

**6. Conclusion**

The problem of an arbitrary truncated Levy flight description by the method of cumulant approach was solved. The set of cumulants of the truncated Levy distribution given the assumption of arbitrary truncation was found. It was shown that a particular dependence of the order of high cumulant coefficient magnitude upon the cumulant order exists, and this dependence is the criterion of belonging to the class of truncated Levy distributions. It was shown that characteristics of truncated Levy flight in the Gaussian regime completely depend upon two increment distribution cumulants, namely, the variance and the kurtosis. The variance and the kurtosis dependences upon the truncation shape were investigated. It was shown that the truncated Levy flight in the Levy regime is described by the complete set of cumulant functions. In addition, the particular property of the truncated Levy flight in the Levy regime one-dimensional probability distribution was found. Namely, the Levy flight *n*th step one-dimensional probability distribution coincides with the distribution, which can be obtained from the initial truncated Levy distribution by changing its spatial scale from $\gamma$ to the value $\gamma_{\textit{eff}} = \gamma \cdot n^{1/\alpha}$ and keeping the spatial scale of truncation, where $\alpha$ is the stability index of the Levy distribution.